\documentstyle[12pt,epsfig]{article}

\newfam\BMath
\font\BMathL=cmmib10 
\font\BMathl=cmmib7
\font\BMathm=cmmib5
\textfont\BMath=\BMathL \scriptfont\BMath=\BMathl
\scriptscriptfont\BMath=\BMathm

\newcommand\cm{{\cal M}}   

\newcommand\dd{\mbox{d}}

\newcommand\G{\Gamma}
\renewcommand\d{\delta}
\newcommand\f{\phi}
\renewcommand\l{\lambda}
\renewcommand\o{\omega}
\newcommand\p{\pi}

\newcommand\be{\begin{equation}}
\newcommand\ee{\end{equation}}
\newcommand\bea{\begin{eqnarray}}
\newcommand\eea{\end{eqnarray}}

\newcommand\bfi{\begin{figure}}
\newcommand\efi{\end{figure}}
\newcommand{\eref}[1]{Eq.~(\ref{#1})}
\newcommand{\fref}[1]{Fig.~\ref{#1}}

\newcommand\intp{\int \frac{\dd^4 p_1}{(2\pi)^4}}
\newcommand\intks{\int \frac{d^3 k}{(2\pi)^3 2 \omega}} 

\newcommand\intpso{\int \frac{d^3 p_1}{(2\pi)^3 2 p^0_1}}
\newcommand\intpst{\int \frac{d^3 p_2}{(2\pi)^3 2 p^0_2}}

\renewcommand\exp{\mbox{\rm exp}} 
\newcommand{\half}{\frac{1}{2}}

\title{\begin{flushright}
{\normalsize NUC-MINN-2001/4-T\\
February 2001 \\}
\end{flushright}
\vspace*{0.3in}
{\bf Two-Loop Self-Energy and Multiple Scattering at Finite Temperature}}
\author{J.I. Kapusta$^1$ and S.M.H. Wong$^2$
\vspace*{0.1in}\\
$^1${\it School of Physics and Astronomy, University of Minnesota}\\ 
{\it Minneapolis, MN 55455} \\
$^2${\it Department of Physics, Ohio State University}\\
{\it Columbus, OH 43210}}

\date{}

\begin{document}

\maketitle

\begin{center}
Abstract
\end{center}

One and two loop self-energies are worked out explicitly for a heavy scalar
field interacting weakly with a light self-interacting scalar field at finite
temperature. The ring/daisy diagrams and a set of necklace diagrams can be 
summed simultaneously. This simple model serves to illustrate the connection 
between multi-loop self-energy diagrams and multiple scattering in a medium.

\vspace*{1.in}

\newpage

The study of how and why the properties of particles change in a thermal medium 
is fascinating and of particular importance in the areas of cosmology, 
astophysics, and high energy nucleus-nucleus collisions.  In this paper we 
investigate a very simple model involving a heavy scalar field with vacuum mass 
$M$ interacting with a much lighter scalar field with vacuum mass $m \ll M$ at 
temperature $T \ll M$.  The lighter field has a quartic self-interaction that is 
stronger than the interaction between the light and heavy fields.  Although the 
model considered here is not realistic it has many properties in common with 
more physically interesting theories, such as the decay of a very heavy boson in 
the early universe or the emission of a very massive virtual photon in high 
energy nuclear collisions.  The simplicity of this model allows a very clear 
mathematical evaluation of the heavy boson self-energy at finite temperature and 
its physical interpretation in terms of multiple scattering in the many-particle 
medium.  One loop self-energies were studied and physically interpreted in a 
beautiful paper by Weldon \cite{art1}. Some examples of two loop 
calculations are \cite{photon,bpy,bnnr,w1} but since those are 
oriented toward real physical applications the resulting formulae 
are necessarily much more involved and oftentimes tedious to reproduce.
We should also like to point out that at low to moderate densities the
self-energy can always be expressed as an integral over the scattering amplitude
for scattering from particles in the medium. Frequently one can use experimental
input of scattering amplitudes to construct the self-energy \cite{f}. 
A formal discussion of this is was presented by Jeon and Ellis \cite{jeon}.

The heavy scalar field is labeled $H$ and the much lighter scalar field is 
labeled $\f$.  The interaction Lagrangian is
\be
{\cal L}_I = gH\f^2 - \l \f^4 \, .
\ee
The cubic coupling $g$ has the dimension of mass while the quartic coupling is 
dimensionless.  It is assumed that $g^2 \ll \l M^2$ in order that we can 
ignore any self-energy diagrams involving an internal $H$ line.  This is 
analogous to the problem of the coupling of a photon to quarks where the 
electromagnetic coupling is much smaller than the strong coupling.

\bfi[b]
\centerline{\epsfig{figure=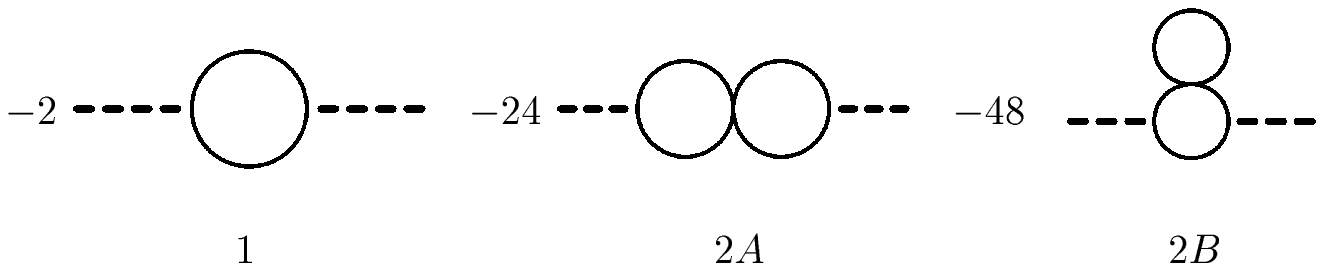,width=11.0cm}}
\caption{The one and two loop contributions to the $H$
self-energy under the condition $g^2 \ll \l M^2$.  The thick dashed line 
is the $H$, the solid line is the $\f$.  There is a factor $g$ at each 
three point vertex and a factor $-\l$ at each four point vertex.}
\label{fig1}
\efi

The resulting one and two loop diagrams for the $H$ are drawn in \fref{fig1}.  
We 
are interested in the limit that the $H$ is at rest with respect to the medium.  
(The interested reader may easily generalize the results that follow.)  We will 
first evaluate the diagrams in Euclidean space and then analytically continue to 
Minkowski space. Denoting the Matsubara frequency of the external $H$ line by 
$\o_n = 2\pi n T$ we can find explicit mathematical expressions for these 
diagrams in the usual way \cite{mybook}.
\bea
\Pi_1 &=& -2g^2 \left[ T \sum_j \int \frac{d^3p}{(2\pi)^3}
\frac{1}{\o_j^2 + p^2 + m^2} \, \frac{1}{(\o_j-\o_n)^2 + p^2 + m^2}
\right] \\
\Pi_{2A} &=& 24 g^2 \l  \left[ T \sum_j \int \frac{d^3p}{(2\pi)^3}
\frac{1}{\o_j^2 + p^2 + m^2} \, \frac{1}{(\o_j-\o_n)^2 + p^2 + m^2}
\right]^2\\
\Pi_{2B} &=& 48 g^2 \l  \left[ T \sum_k \int \frac{d^3q}{(2\pi)^3}
\frac{1}{\o_k^2 + q^2 + m^2}\right] \nonumber \\
&&  \left[ T \sum_j \int \frac{d^3p}{(2\pi)^3}
\frac{1}{\o_j^2 + p^2 + m^2}
 \left( \frac{1}{(\o_j-\o_n)^2 + p^2 + m^2} \right)^2 \right]
\eea
These summations may be evaluated directly or by the use of contour integration.

The simplest summation is
\be
F(m,T) = T \sum_j \int \frac{d^3p}{(2\pi)^3}
\frac{1}{\o_j^2 + p^2 + m^2} = \int \frac{d^3p}{(2\pi)^3}
\frac{1}{\o} \left[ n_{\rm BE}(\o/T) + \frac{1}{2} \right] 
\ee
where $\o = \sqrt{p^2 + m^2}$ and $ n_{\rm BE}(x)=1/(e^x-1)$ is the
Bose-Einstein occupation number.  The next simplest is
\bea
G(\o_n,m,T) &=& T \sum_j \int \frac{d^3p}{(2\pi)^3}
\frac{1}{\o_j^2 + p^2 + m^2} \, \frac{1}{(\o_j-\o_n)^2 + p^2 + m^2}
\nonumber \\
&=& \int \frac{d^3p}{(2\pi)^3} \frac{1}{\o} \, 
\frac{1}{4\o^2+\o_n^2} \left[ 2n_{\rm BE}(\o/T) + 1 \right] 
\eea
The final summation needed is
\be
T \sum_j \int \frac{d^3p}{(2\pi)^3}
\frac{1}{\o_j^2 + p^2 + m^2}
 \left( \frac{1}{(\o_j-\o_n)^2 + p^2 + m^2} \right)^2 
= -\frac{1}{2} \frac{\partial}{\partial m^2} G(\o_n,m,T) \, .
\ee
We can express the one and two loop self-energies in terms of these functions.
\bea
\Pi_1 &=& -2g^2 G(\o_n,m,T) \\
\Pi_{2A} &=& 24 \l g^2 G^2(\o_n,m,T) \\
\Pi_{2B} &=& -24 \l g^2 F(m,T)
\frac{\partial}{\partial m^2} G(\o_n,m,T)
\eea
In Euclidean space the self-energy is real.  Note that both the one and two loop 
contributions can be expressed in terms of one dimensional integrals.

In order to obtain physical quantities, such as the dispersion relation, 
scattering, decay, and production rates, we must analytically continue to 
Minkowski space.  The retarded self-energy is obtained by the replacement
$\o_n \rightarrow -iM+\epsilon$ with $M>0$ and $\epsilon \rightarrow 0^+$.  
Then
\bea
Im \, G &=& \frac{1}{16\pi} \sqrt{1-\frac{4m^2}{M^2}}
\left[ \left( n_{\rm BE}(M/2T)+1\right)^2 - n_{\rm BE}^2(M/2T) \right] \\
Re\, G &=& \frac{1}{2\pi^2} P.V. \int_0^{\infty} \frac{dp p^2}{\o}
\frac{1}{4\o^2-M^2} \left[ 2n_{\rm BE}(M/2T)+1\right] \, .
\eea
Here $P.V.$ stands for principle value.  The 1 following the Bose-Einstein 
occupancy number in the real part above represents the vacuum contribution.  To 
really evaluate it we should go back to a four dimensional integration over 
momentum in Euclidean space and place an upper cutoff on the integration, or 
else analytically continue to $4+\epsilon$ dimensions; in either case 
renormalization would proceed as usual.  See the appendix.  The term
$(n_{\rm BE}+1)^2$ in the 
imaginary part represents the decay $H \rightarrow \f \f$ with the
Bose-Einstein enhancement factor in the final state; the term $n_{\rm BE}^2$ 
represents the production $\f \f \rightarrow H$.

The real part of $\Pi$ represents a shift in the mass squared of the $H$ and the 
imaginary part represents the rate of its decay, scattering, or production.  It 
is instructive to consider the limit $m \ll T \ll M$.
\bea
Re \, G &\approx & - \frac{T^2}{6M^2} \\
F &\approx & \frac{T^2}{12} + {\rm vacuum}
\eea
The imaginary part of the $H$ self-energy at the one and two loop order now 
takes a very simple form.
\be
Im \, \Pi = -\frac{g^2}{8\pi}   
\left[ \left( n_{\rm BE}+1\right)^2 - n_{\rm BE}^2 \right]
\left\{ 1 + (4-2) \l \frac{T^2}{M^2} + ... \right\}
\label{imPI}
\ee
This formula may be interpreted thusly.  The overall sign results from the 
convention used for the self-energy.  A minus sign indicates loss or decay of 
the $H$, a plus sign indicates gain or production.  The factor
$\left( n_{\rm BE}+1\right)^2$ shows the Bose-Einstein enhancement of the final 
state, the factor $n_{\rm BE}^2$ shows that the probability of production is 
proportional to the probability to find two $\f$ of the requisite energy in 
the initial state.  (The $n_{\rm BE}$ are of course evaluated at the energy 
$M/2$.)  The correction to order $\l$ has contributions from the diagrams 
$\Pi_{2A}$ giving the factor 4 and from $\Pi_{2B}$ giving the factor $-2$.  Is 
there a simple interpretation of these results?

The imaginary part can be obtained by cutting a diagram in half.  Cutting the 
one loop diagram, as in \fref{fig2}, illustrates the physical process of decay 
of 
the $H$ into $\f \f$.  (For definiteness we focus on the absorption or decay 
processes.  Reversing the in and out states gives the production rate.)
\bfi[t]
\centerline{\epsfig{figure=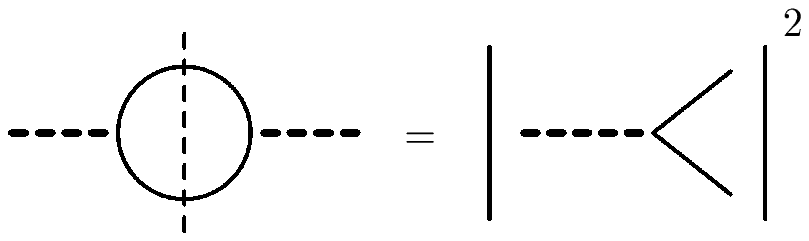,width=7.0cm}}
\caption{Cutting the one loop self-energy diagram.}
\label{fig2}
\efi

Cutting the diagram for $\Pi_{2A}$ does not yield the square of an amplitude.  
Instead it illustrates an interference between an amplitude of order zero  
and an amplitude of order $\l$.  See \fref{fig3}. Note that there 
should be a spectator $\f$ with energy $\o$ in the simpler half of 
the interference graphs as shown in previous work \cite{kw,w3}. 
We have not included it here to reduce the number of sub-figures 
within each figure that need to be drawn.
The same also applies to \fref{fig4} below. The loop that is not 
cut involves a Matsubara sum. This can be expressed as a vacuum loop plus a 
finite temperature piece.  The former is just the order $\l$ correction 
to the vertex coupling the $H$ to $\f \f$.  The latter can be viewed as 
the absorption of a $\f$ from the heat bath with energy $\o$ and 
momentum ${\bf p}$ and the emission of a $\f$ into the heat bath with the 
same energy and momentum. The initial and final states are the same as in 
the amplitude of lower order in $\l$ provided a spectator is included 
in the simpler half of the interference graph as mentioned above, hence 
the reason that it can interfere. There are two diagrams with a single 
intermediate off-shell $\f$.  The kinematics are such that one of them 
has propagator $1/q_+^2$ with $q_+^2 = M(M+2\o)$, and the other has 
propagator $1/q_-^2$ with $q_-^2 = M(M-2\o)$.  Adding these two 
propagators gives $2/(M^2-4\o^2)$.  This is precisely the origin of 
the mathematical expressions displayed in eqs. (6) and (12) above.
\bfi[h]
\centerline{\epsfig{figure=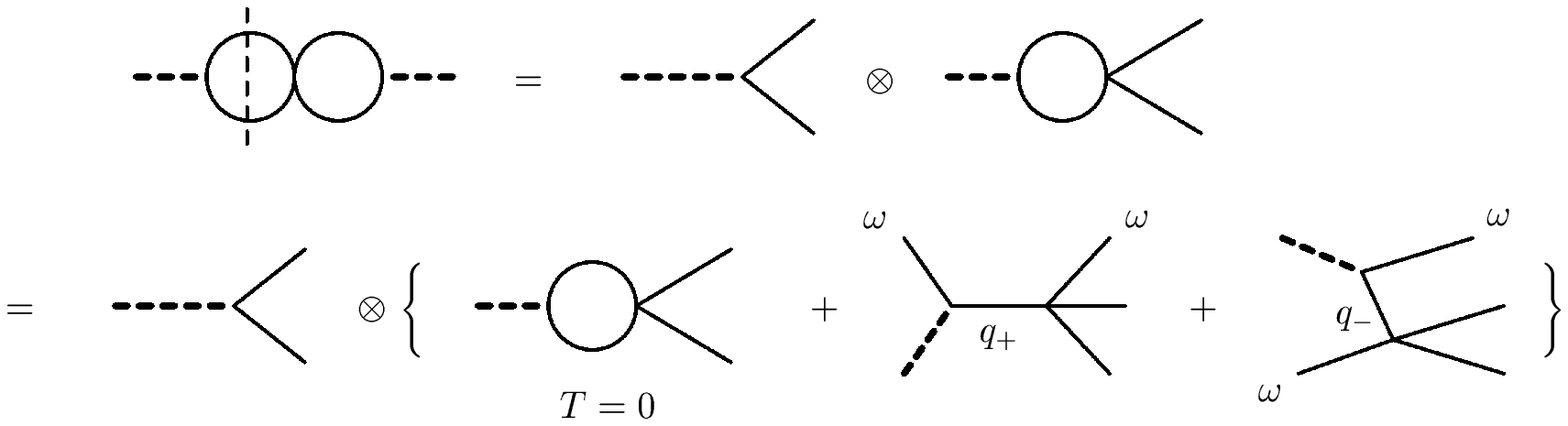,width=13.0cm}}
\caption{Cutting the two loop diagram for $\Pi_{2A}$ results in an
interference between two amplitudes of different order in $\l$.  A loop
involves a Matsubara sum unless otherwise indicated as a vacuum loop. 
The $q_+$ and $q_-$ are the four-momenta of intermediate states.
Combinatoric factors are not displayed.}
\label{fig3}
\efi

\newpage
 
\bfi
\centerline{\epsfig{figure=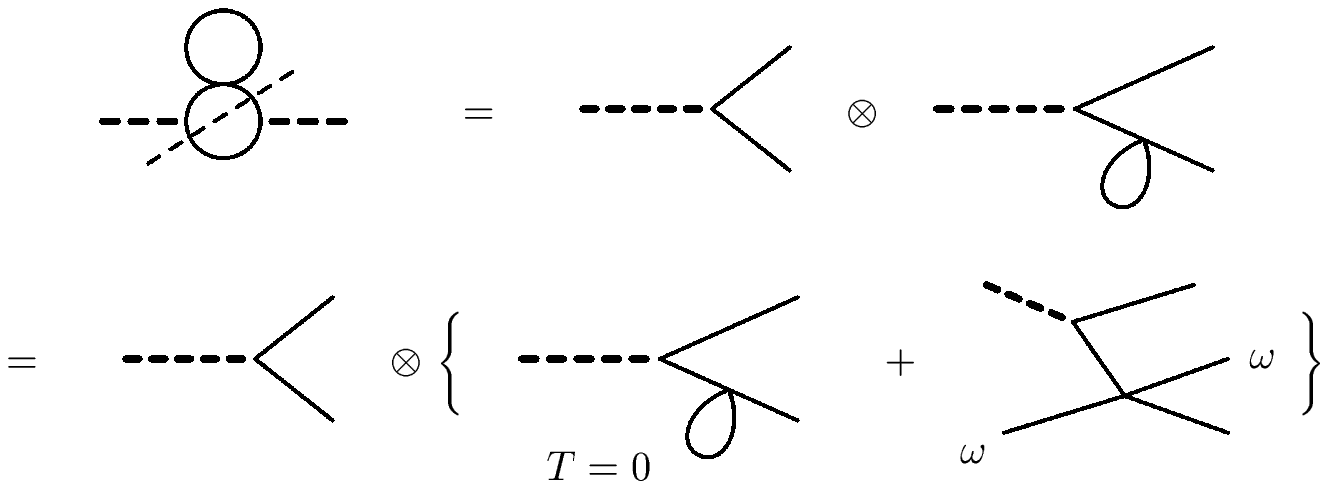,width=12.0cm}}
\caption{Cutting the two loop diagram for $\Pi_{2B}$ is just the second 
term in the Taylor series expansion for the one loop diagram shown in figure 2 
with the bare $\f$ propagator replaced by the dressed one.  In effect the in 
and out $\f$ lines should be dressed to represent collective excitations in 
the thermal medium rather than vacuum particles.}
\label{fig4}
\efi

Cutting the diagram for $\Pi_{2B}$ is shown in \fref{fig4}.  Again there 
is an interference between amplitudes of different order in $\l$.  One 
is tempted to interpret this as scattering of a particle from the heat bath 
with one of the outgoing $\f$ from the $H$ decay.  However, because there 
is an intermediate line on-shell the diagram is not well-defined mathematically 
(one should also read the paragraph near the end of the paper which
discusses how in the imaginary-time formalism the contour integral allows 
one to avoid this problem altogether). This should be no surprise: one does 
not dress external lines when evaluating Feynman diagrams for scattering 
or decay. Indeed, this diagram really arises from an expansion of $\Pi$ in 
terms of a dressed $\f$ propagator and bare vertices rather than an 
expansion in terms of a bare $\f$ propagator and bare vertices. The one 
loop contribution to the self-energy of the $\f$ is momentum and frequency 
independent and is simply given by $12 \l F(m,T)$.  The dressed propagator 
is $1/(\o_j^2 + p^2 + m_T^2)$ where $m_T^2 = m^2 + 12 \l F(m,T)$.  One 
sees very clearly that $\Pi_{2B}$ is the second term in a Taylor series, the 
first term being $\Pi_1$.  All diagrams of this type can easily be summed by 
making the replacement $\Pi_1(\o_n,m,T) \rightarrow \Pi_1(\o_n,m_T,T)$.  
Physically what is happening is that the $H$ is decaying into collective 
excitations with the quantum numbers of the $\f$ rather than into vacuum-like 
$\f$ excitations.  In the context of gauge theories this goes by the name of 
{\it hard thermal loops} \cite{resum}.  It is left as an exercise for the 
reader to show that the combinatoric factor for each diagram in this class is 
exactly reproduced by this selective resummation.  Such perturbative diagrams 
go by the name of ring or daisy diagrams.  One can go even further by summing 
super ring or super daisy diagrams by solving the integral equation 
\be
\overline{m}^2_T = m^2 + 12 \l F(\overline{m}_T,T)
\ee
for the effective mass $\overline{m}_T$.  Examples of ring/daisy and super 
ring/daisy diagrams are shown in \fref{fig5} (a) and (b) respectively.

\newpage

\bfi[t]
\centerline{\epsfig{figure=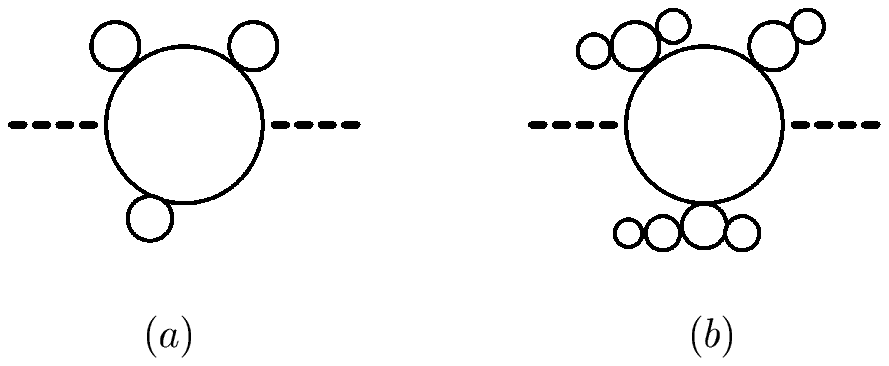,width=9.0cm}}
\caption{(a) an example of a ring/daisy diagram, and (b) an example of a super 
ring/daisy diagram.}
\label{fig5}
\efi

There is another class of disgrams that are easily summable.  We refer to these 
as necklace diagrams.  They have exactly the same topological form as $\Pi_{2A}$ 
but with $N$ loops instead of 2.  The overall combinatoric factor is $-12^N/6$.  
The expression for the one with $N$ loops is
\be
\Pi_N^{\rm necklace} = -\frac{12^N}{6} g^2 (-\l)^{N-1} G^N(\o_n,m,T) 
\, .
\ee
This is a geometric series readily summed to give
\be
\Pi_{\rm necklace} = - \frac{2g^2 G(\o_n,m,T)}{1+12 \l G(\o_n,m,T)} 
\, .
\ee
Cutting the $N=3$ necklace, for example, leads to the diagrams shown in 
\fref{fig6} and \fref{fig7}. At first one think that summing the 
necklaces amounts to a temperature coupling for $H\f\f$, namely, 
$g^2(T)$.  This is not a good way to think because not only would this 
effective coupling have a dependence on the external Matsubara frequency 
$\o_n$ but it will become complex when analytically continuing to 
Minkowski space. 

\bfi[b]
\centerline{\epsfig{figure=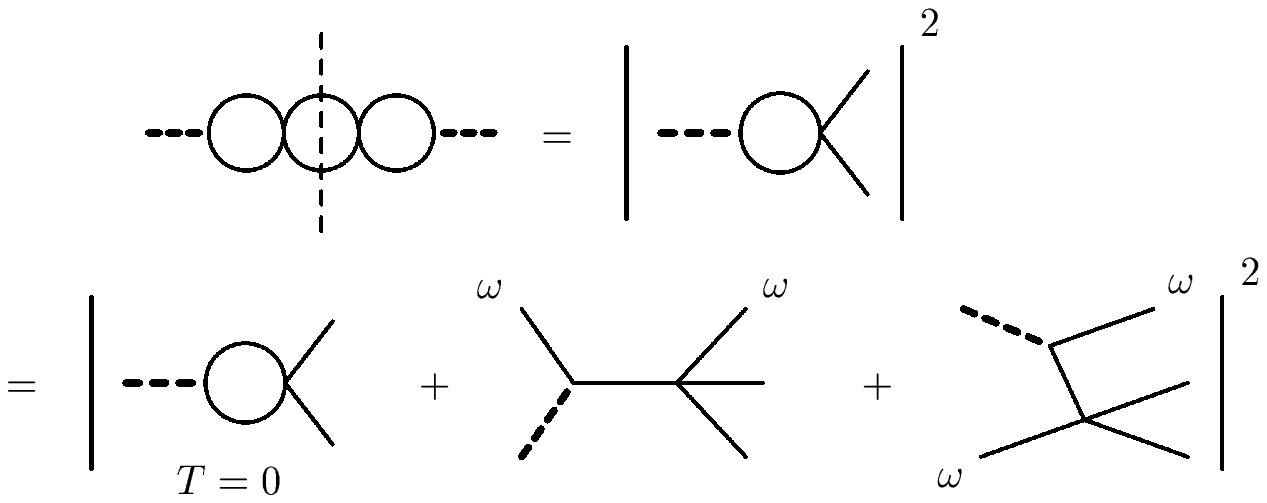,width=11.0cm}}
\caption{One way of cutting the $N=3$ necklace diagram.
Combinatoric factors are not displayed.}
\label{fig6}
\efi

\newpage

\bfi[t]
\centerline{\epsfig{figure=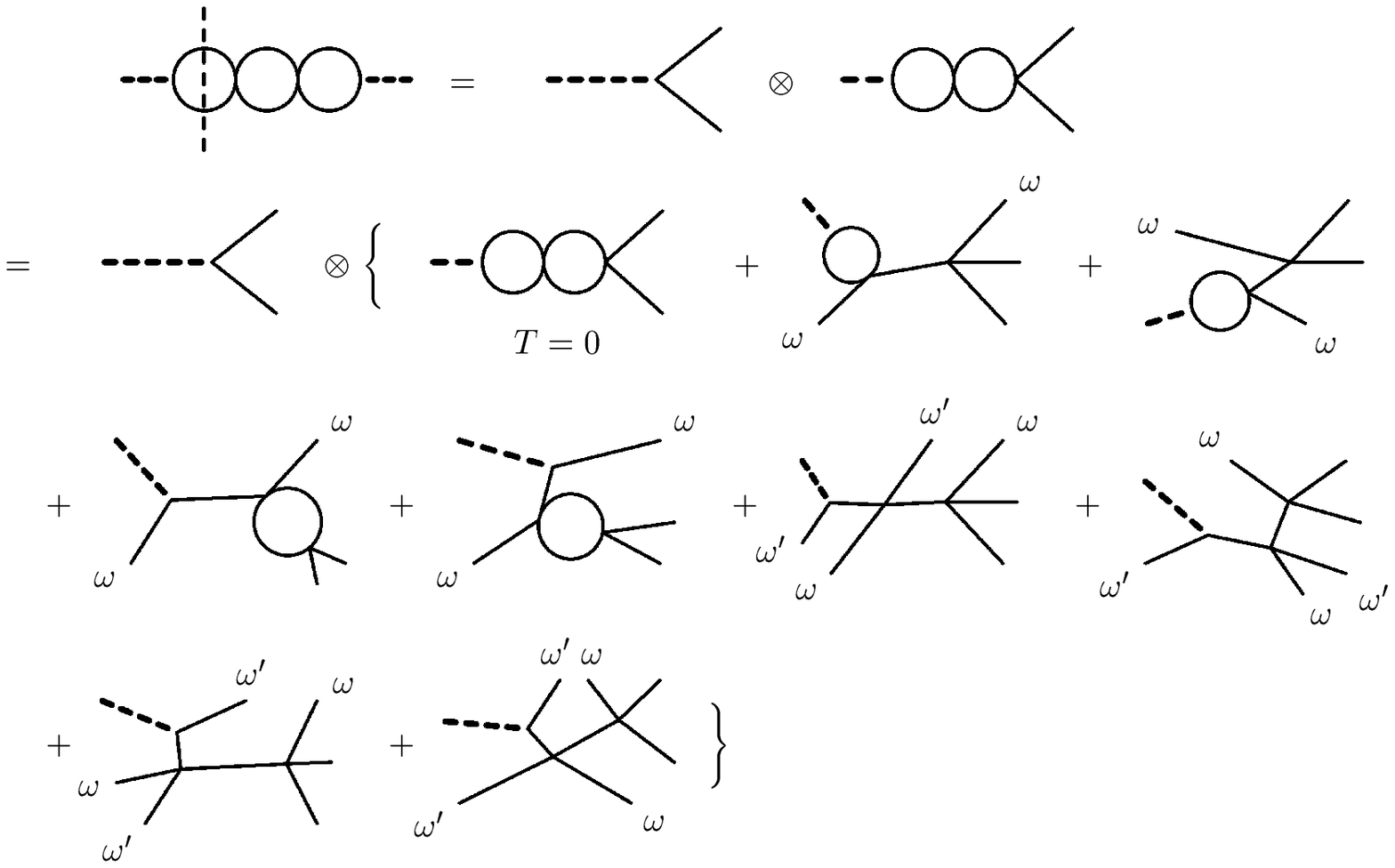,width=13.0cm}}
\caption{Another way of cutting the $N=3$ necklace diagram.
All the loops between the brackets are evaluated at T=0.
Combinatoric factors are not displayed.}
\label{fig7}
\efi

One can further sum all diagrams of the ring and necklace types simultaneously 
simply by replacing $m$ in the necklace formula with $\overline{m}_T$.
\be
\Pi_{\rm ring + necklace} = - \frac{2g^2 G(\o_n,\overline{m}_T,T)}
{1+12 \l G(\o_n,\overline{m}_T,T)} \, .
\ee
Neglecting the denominator, and expanding the numerator to zeroth and first 
order in $\l$ reproduces $\Pi_1$ and $\Pi_{2B}$, respectively.  Keeping the 
numerator to zeroth order and expanding the denominator to first order 
reproduces $\Pi_{2A}$.

We can also calculate the imaginary part for \fref{fig1} from the scattering 
amplitudes shown in Figs. 2-4.  This procedure was used in \cite{kw} and 
described in more detail in \cite{w3}. Rather than calculating the loop diagrams 
in Euclidean space, analytically continuing to Minkowski space, and then 
finding the imaginary part, this procedure uses the scattering amplitudes as 
input and does a thermal averaging over particles in the heat bath. In this 
procedure, every line carrying a distinct 4-momentum is integrated over phase 
space with the appropriate thermal weight.  The integrations are of course 
constrained by a global 4-momentum conserving delta-function.  The amplitudes 
are evaluated at tree level, except for the explicit vacuum loops.  There are 
no finite temperature loops.  The amplitudes are computed according to 
conventional 
vacuum field theory.  In those cases where an outgoing line is in effect dressed 
by the scattering of a particle from the heat bath, as in Fig. 4, then the 
amplitude 
will have to be dealt with more carefully; see below and also ref. \cite{w3} for 
a general discussion. 

Applying the scattering amplitude method to the current calculation 
of the decay of the heavy scalar field $H$ at rest, one starts directly 
with the interference graphs on the second line in \fref{fig3}. A phase 
space integration measure is assigned to each independent incoming 
and outgoing line except, of course, the $H$ line. The two lines carrying
the energy $\o$ will have only one and {\em not} two integral 
measures because the same 4-momentum flows through them. They come from the heat 
bath and go back into it with the same momentum, therefore they are not 
constrained by the overall energy-momentum conservation. The contribution 
from \fref{fig3} to the decay rate $\G$ can be written as
\bea 2 E \G_{ \mbox{\scriptsize (\fref{fig3})} } &=& \frac{1}{2}
     \intpso \intpst \; (2\p)^4 \d^{(4)} (P-p_1-p_2) 
     \nonumber \\
     & & \times 
     \Big (1+n_{\rm BE}(p_1^0/T) \Big )\Big (1+n_{\rm BE}(p_2^0/T) \Big )   
     \Big (  \cm_{\mbox{\scriptsize tree}}^* \otimes 
             \cm_{ \mbox{\scriptsize (\fref{fig3})} }
           + \mbox{h. c.} \Big )  \; .
\label{rf3}
\eea 
The factor of minus one-half is a symmetry factor. Here $P = (E,{\bf P})$ is
the 4-momentum of the $H$. In this paper, though, we specialize to the $H$ at 
rest so that ${\bf P}=0$ and $P^0=M$. For the simpler half of the interference
graph, $\cm$ is just the tree amplitude \be \cm_{\mbox{\scriptsize
    tree}} =  2 g \; .
\label{mtr}
\ee
The other half is more complicated. 
\bea \cm_{ \mbox{\scriptsize (\fref{fig3})} } &=& 2!\times 4!\times g(-\l) 
     \intks \Big (\half +n_{\rm BE}(\o/T) \Big ) \; 
     \bigg ( - \frac{1}{q^2_+} - \frac{1}{q^2_-} \bigg )  \nonumber \\
     & = & \frac{4 g \l T^2}{M^2} + \{T=0\; \mbox{part}\} + {\cal O}(1/M^4) 
     \;\; .
\label{mf3}
\eea 
The factors 2! and 4! come from the $H\phi^2$ vertex and 
the $\phi^4$ vertex, respectively. The $(1/2+n_{\rm BE})$ is
the distribution assigned to boson lines which correspond to absorption 
and emission of the same 4-momenta within the same Feynman graph
according to one of the rules given in \cite{w3}. Its origin is really just 
the same as Eq. (5).  The denominators
$q^2_+$ and $q^2_-$ introduced earlier come out automatically
from calculating the last two graphs in \fref{fig3}.  They have a minus
sign because of our conventions, namely, the propagator is positive
in Euclidean space \cite{mybook}.  With \eref{mtr} and \eref{mf3},
\eref{rf3} can easily be evaluated to give 
\be  2 M \G_{ \mbox{\scriptsize (\fref{fig3})} } 
     = \frac{\l g^2 T^2}{\p M^2} + \{T=0\; \mbox{part}\} 
      +\{\mbox{non-leading terms}\} \; .
\ee

The other contribution to the $H$ decay comes from the processes 
shown in \fref{fig4}. This time there is an intermediate line which
is on-shell because it carries the same 4-momentum as an outgoing line.
For that line, with 4-momentum $p_1$, the phase space integral measure 
must be written out as a 4-momentum integral measure because 
there is a differentiation of the mass-shell constraining delta-function.
The intermediate on-shell line is mathematically 
absorbed into this differentiation of the mass-shell constraint and 
does not appear anywhere else in the calculation.  In terms of the 
self-energy loop diagrams shown in Fig. 1, this derivative of a Dirac 
delta-function has its origins in a double pole arising from a self-energy 
insertion.  The contribution from \fref{fig4} can be written as 
\bea 2 M \G_{ \mbox{\scriptsize (\fref{fig4})} } &=& \frac{1}{2}\times 2
     \intp  \intpst \; 2\p \d^{'(+)}(p^2_1) \; 
     (2\p)^4 \d^{(4)} (P-p_1-p_2)                        \nonumber \\
     & & \times 
     \Big (1+n_{\rm BE}(p_1^0/T) \Big )\Big (1+n_{\rm BE}(p_2^0/T) \Big )   
     \Big (  \cm_{\mbox{\scriptsize tree}}^* \otimes 
             \cm_{ \mbox{\scriptsize (\fref{fig4})} } 
           + \mbox{h. c.} \Big )  \; .
\label{rf4}
\eea 
The factor of two here is for the emission and absorption to occur on either
of the two outgoing lines. 
$\cm_{\mbox{\scriptsize tree}}$ is the same as above in \eref{mtr}.
The symbol $\d^{'(+)}$ means to differentiate the Dirac delta-function with
respect to its argument, taking the energy positive. 
The remaining half of the amplitude which contains the emission and 
absorption is 
\bea \cm_{ \mbox{\scriptsize (\fref{fig4})} } &=&2!\times 4!\times g(-\l) 
     \intks \;\Big (\half +n_{\rm BE}(\o/T) \Big ) \nonumber \\
     & = & - 2 g\l T^2 + \{T=0\; \mbox{part}\} 
     \;\; .
\label{mf4} 
\eea 
The factors of 2! and 4! have the same origin as
before, and the emission/absorption is associated with a distribution 
of $(1/2+n_{\rm BE})$. The integration in \eref{rf4} has to be
evaluated with some care in order to manipulate correctly the 
differentiation of the mass-shell constraint.  The result is
\be 2 M \G_{ \mbox{\scriptsize (\fref{fig4})} } 
    = -\frac{\l g^2 T^2}{2\p M^2} + \{T=0\; \mbox{part}\} 
      +\{\mbox{non-leading terms}\}  \; .
\ee 
So we have 
\be 2 M \Big ( \G_{ \mbox{\scriptsize (\fref{fig3})} } 
              +\G_{ \mbox{\scriptsize (\fref{fig4})} } \Big ) 
    \simeq   (2-1)\;\frac{\l g^2 T^2}{2 \p M^2} 
      + \{T=0\; \mbox{part}\}  \; . 
\ee
The result from \eref{imPI} can be recovered with 
\be \G = \G_{ \mbox{\scriptsize (\fref{fig2})} } 
        +\G_{ \mbox{\scriptsize (\fref{fig3})} }
        +\G_{ \mbox{\scriptsize (\fref{fig4})} } 
\ee 
using the identification
\be 2 M \G \; \Big (\exp (-M/T)-1 \Big ) = 2\; Im\; \Pi   \; .
\ee 

One point in the calculation that should be elaborated
concerns self-energy insertions.  In terms of scattering amplitudes it 
amounts to dressing an external line, thereby causing a singularity on 
an internal line.  The usual Feynman rules for scattering theory say 
that external lines should not be dressed but should be taken into 
account in the phase space. (In our example, this is the intermediate 
line in \fref{fig4}.) This usually leads to the need to
include at least part of the thermal self-energy in the field, and 
therefore changes the mass-shell of the particle.  This was shown quite 
elegantly by Weldon \cite{art2} for fermions. 
Donoghue and Holstein have done the same but in a Lorentz 
non-covariant manner \cite{JB}.
When calculating loops in the imaginary-time formalism, there is a 
rigorous solution to this problem. It is well-known that the 
contour integral method has a well-defined way of dealing with 
multi-pole singularities. The residue of the multi-pole is found by 
differentiating the non-pole part of the integrand by one power less than 
that of the order of the multi-pole. So, if one insists on no resummation
of any sort, including those of super ring/daisy diagrams mentioned
above, or even partial resummation, such as those of Weldon, Donoghue and 
Holstein, then the contour method is a good solution to the above 
problem. This method was also used in \cite{kw,w3}. 

In conclusion we have accomplished several goals.  First, we have explicitly 
evaluated the two loop self-energy for a massive boson weakly interacting with 
lighter bosons in thermal equilibrium among themselves.  This was done first in 
the imaginary time formalism of Matsubara.  There is no ambiguity in 
analytically continuing to real time or real energies, as first proven by Baym 
and Mermin 40 years ago \cite{BM}.  Second, a physical interpretation was found 
for all 
diagrams considered.  Some of these involve Bose-Einstein enhancement of 
strongly interacting particles in the final state of the heavy boson decay.  
Some of them involve absorption and emission of particles from the heat bath 
with the same energy and momentum.  Such processes may interfere with processes 
of other order in the stronger coupling $\l$.  Third, it was seen how 
important it is to expand in terms of resummed or interacting propagators. This 
indicates the importance of identifying the collective excitations of the 
system.  These are all features that must be understood and taken into account 
when considering the kinetic theory of quark-gluon plasma in heavy ion 
collisions or in the early universe, for example.  Then the role of quarks and 
gluons are analogous to the $\f$ bosons, and the role of the $H$ is analogous 
to the electromagnetic field.

\section*{Acknowledgements}

This work was supported by the US Department of Energy under grant
DE-FG02-87ER40328.  S.W. would like to thank the Physics Department at 
Ohio State University for support.

\newpage

\section*{Appendix}

In this appendix we compute the vacuum self-energy of the H boson to
one and two loop orders.  The diagrams are those shown in Fig. 1.  
We use dimensional regularization (see, for example,
chapter 4 of \cite{Ramond})
and take the mass $m$ of the $\phi$ boson to be zero for simplicity
of presentation.  The two integrals which are needed are the $F$ and
$G$ as displayed in eqs. (2-10).  In Euclidean space:
\begin{equation}
F_{\rm vac} = (\mu^2)^{2-\omega} \int \frac{d^{2\omega}l}{(2\pi)^{2\omega}}
\frac{1}{l^2}
\end{equation}
and
\begin{equation}
G_{\rm vac} = (\mu^2)^{2-\omega} \int \frac{d^{2\omega}l}{(2\pi)^{2\omega}}
\frac{1}{l^2} \frac{1}{(l-P)^2} \, .
\end{equation}
Here $l$ is the loop momentum and $P$ is the external momentum, both Euclidean.
The number of dimensions is $2\omega$ with $2-\omega = \epsilon
\rightarrow 0^+$, and $\mu$ is the dimensional parameter which is arbitrary.
Dimensional regularization gives $ F_{\rm vac}=0$.

Using the Feynman parametrization one arrives at
\begin{eqnarray}
G_{\rm vac} &=& \frac{\Gamma(\epsilon)}{(4\pi)^2}
\left(\frac{4\pi\mu^2}{P^2}\right)^{\epsilon}
\int_0^1 \frac{dx}{[x(1-x)]^{\epsilon}}
= \frac{1}{(4\pi)^2} \left\{ \frac{1}{\epsilon} 
+ \left[2-\gamma_E + \ln\left(\frac{4\pi\mu^2}{P^2}\right)
\right] \right. \nonumber \\
&+& \left. \left[ 4-\frac{\pi^2}{12} -2\gamma_E
+\frac{1}{2}\gamma_E^2 +(2-\gamma_E) \ln\left(\frac{4\pi\mu^2}{P^2}\right)
+\frac{1}{2} \ln^2\left(\frac{4\pi\mu^2}{P^2}\right) \right]\epsilon
\right. \nonumber \\
&+& \left. {\cal O}(\epsilon^2) \right\} \, .
\end{eqnarray}
Analytic continuation to Minkowski space involves the replacements
$P^2 \rightarrow -P^2$ and $\ln\left(4\pi\mu^2/P^2\right) 
\rightarrow \ln\left(4\pi\mu^2/P^2\right) +i\pi$.  This results
in $G_{\rm vac}$ having both real and imaginary parts.

The imaginary part of the $H$ vacuum self-energy can easily be
found from the above results in conjunction with eqs. (8-10).  At
one loop order
\begin{equation}
Im \, \Pi_1^{\rm vac} = -\frac{g^2}{8\pi}
\end{equation}
and at two loop order
\begin{equation}
Im \, \Pi_2^{\rm vac} = \frac{3\lambda g^2}{16\pi^3}
\left\{ \frac{1}{\epsilon} + 4 -2\gamma_E + 
2\ln\left(\frac{4\pi\mu^2}{P^2}\right) \right\} \, .
\end{equation}
This needs a counterterm added to the Lagrangian to render it finite
in the limit $\epsilon \rightarrow 0$.
We have the freedom to choose the counterterm such
that the order $\lambda$ correction vanishes on the mass-shell.
If we make this choice then
\begin{equation}
Im \, \Pi_2^{\rm vac} = \frac{3\lambda g^2}{8\pi^3}
\ln\left(\frac{M^2}{P^2}\right) \, ,
\end{equation}
where $M$ is the physical mass of the $H$ boson.  In effect this defines
the normalization point of the coupling $g$ of the $H$ to two $\phi$ mesons.
The sum of one and two loops is
\begin{equation}
Im \, \Pi_{\rm vac} = -\frac{g^2(M^2)}{8\pi}
\left[1- \frac{3\lambda}{\pi^2}
\ln\left(\frac{M^2}{P^2}\right) \right]
\approx -\frac{1}{8\pi} \frac{g^2(M^2)}{1+\frac{3\lambda}{\pi^2}
\ln\left(\frac{M^2}{P^2}\right)}
=  -\frac{g^2(P^2)}{8\pi} \, .
\end{equation}
This involves the renormalization group running coupling.
It displays infrared freedom, $g^2(P^2) \rightarrow 0$ as $P^2
\rightarrow 0$, and the Landau pole at $P^2 = M^2 \exp(\pi^2/3\lambda)$.

The above results can also be obtained from the vacuum diagrams shown
in Figs. 2 to 4.  The methods for doing this are well known: see, for
example, chapter 2 of \cite{Field}.  This is left as an exercise for the reader.

\end{document}